# Investigation of the Alzofon weight reduction experiment using NMR spectroscopy

**Willy Stark[1*]**
*Technische Universität Dresden, Dresden, 01069, Germany*

**Hans-Joachim Grafe[2]**
*Leibniz-Institute for Solid State- and Materials Research IFW Dresden, Dresden, 01069, Germany*

**Martin Tajmar[3]**
*Technische Universität Dresden, Dresden, 01069, Germany*

**Abstract**

Interstellar travel requires propulsion systems beyond present possibilities and scientists search for new technologies and breakthrough concepts in physics. Frederick E. Alzofon came up with an alternative idea on the origin of the gravitational field. He claimed that the gravitational force arises from the interaction of subatomic particles and that the gravitational force between the particles of a system decreases during a process of retrograding from an ordered system into a disordered system. To proof his assumption, Alzofon used pulsed dynamic nuclear polarization to create such an order of particles within a test object and claims to have measured distinct changes in weight. This paper describes an experiment to investigate Alzofon's assumptions with the use of NMR spectroscopy for nuclear spin polarization. It could be shown that with certain measurement parameters, an apparent change in weight was indeed measured, but finally this turned out to be a temperature effect.

**Nomenclature**

| | | | |
|---|---|---|---|
| $b$ | Width | $\lambda$ | London penetration depth |
| $B_1$ | Magnetic flux density | $\lambda_C$ | Compton wavelength |
| $c$ | Speed of light | $m$ | Mass |
| $C_{match}$ | Matching capacitor | $\mu$ | Permeability |
| $C_{tune}$ | Tuning capacitor | $\mu_0$ | Magnetic constant |
| $\delta$ | Skin depth | $n$ | Refractive index |
| $E$ | Young's modulus | $\eta$ | Ratio of excited specimen volume |
| $f$ | Frequency | $p$ | Pressure |
| $\gamma$ | Gyromagnetic ratio of the nucleus | $P$ | Power |
| $H_0$ | External Magnetic field strength | $\Pi$ | Ratio of specimen volume affected by NSP |
| $h$ | Height | $Q$ | Quality factor |
| $h_P$ | Planck's constant | $R$ | Resistance |
| $I$ | Nuclear spin | $R_s$ | Specific gas constant |
| $I_y$ | Moment of inertia | $\rho$ | Density |
| $\langle\hat{I}_z\rangle$ | expected value of NSP | $\rho_R$ | Electrical resistivity |
| $k$ | Conversion factor | $T$ | Temperature |
| $k_B$ | Boltzmann constant | $T1$ | Spin–lattice relaxation time |
| $l$ | Length | $V$ | Volume |
| $L$ | Coil | $\omega$ | Angular frequency |

[1*] Corresponding Author, Research Associate, Institute of Aerospace Engineering
E-mail: Willy.Stark@tu-dresden.de
[2] Group Leader Magnetic Properties, Institute for Solid State Research,
E-mail: H.Grafe@ifw-dresden.de
[3] Institute Director, Professor and Head of Space Systems Chair, Institute of Aerospace Engineering,
E-mail: Martin.Tajmar@tu-dresden.de

**Abbreviations**

| | |
|---|---|
| CF | Conversion factor |
| IFW | Leibniz Institute for Solid State and Materials Research Dresden |
| LIF | Laser Interferometer |
| NMR | Nuclear magnetic resonance |
| NSP | Nuclear spin polarization |
| PTFE | Polytetrafluoroethylene |
| RE | Refractive error |
| RT | Room temperature |
| YBCO | Yttrium barium copper oxide |

**Introduction**

In 1981 Frederick E. Alzofon published a paper about the possibilities to counter gravity with present technologies [1]. Based on Einstein's mass-energy equivalence, Alzofon states that the elementary particles are made up of electromagnetic energy or field. "Matter is condensed electromagnetic energy […] which is increasingly dense towards the center and less towards the outside until it shades off into a field – the gravitational field" [2, p.211]. This internal energy is bounded within a small space. As the mass of the particle is constant, the stability of the particle is expressed as the Compton wavelength $\lambda_C$, which therefore is also constant:

$$\lambda_C = \frac{h}{m\,c} \quad (1)$$

where $h$ is the Planck's constant, $m$ is the mass of the particle and $c$ is the speed of light. Alzofon states, that the Compton wavelength describes the extent of the particle's internal energy. In the presence of another particle, the distributions of energy described by this parameter overlap and the two particles share some energy or mass together. Therefore, the single particle gained some mass, decreases its Compton wavelength and changes its conditions for stability. To remain stable, the mass-energy distribution is compressed into a smaller volume with the diameter given by the new Compton wavelength. This draws the second particle closer to the central location of the first particle and causes a reaction which is exerted as an attractive force by the second particle. This reaction should also apply to the reverse case and create attractive forces between both particles – the gravitational force.

To proof this assumption, Alzofon was looking for a way to show that the gravitational force can be reduced. His idea was to convert the internal disordered system of particles of an object into an externally ordered system. He assumed that the energy that it takes the system to retrograde into disorder decreases the gravitational force between the particles and its surrounding. In order to exert such an external order on the particles, Alzofon suggested pulsed dynamic nuclear polarization applied to paramagnetic atoms in a strong magnetic field. Usually, the nuclear spins inside the atoms of a material are randomly orientated within all orientations but can be aligned by applying a strong magnetic field. The atoms can be excited by pulsed frequencies of the same frequency as the Larmor frequency of the atom. The Larmor frequency is the procession of the nuclear spins. By exposing the atoms to this frequency, the electrons start to process even faster and get energized, which causes the electrons and the nucleus to orientate along with the magnetic field. Switching off the pulsed frequencies leaves all nuclear spins orientated in the same direction, but they immediately start to disorientate. Alzofon assumed that the gravitational fields of the surroundings interact with the orientated nuclei and cause the disorientation of the nuclei. He stated that this will temporarily weaken the gravitational field in the surroundings and therefore lead to a reduction of the Earth's gravitational force on the specimen which causes a temporarily loss of weight. More details of Alzofon's assumptions are described elsewhere [1, 2].

His experiments to proof the predicted effect took place in 1994. Alzofon chose aluminum powder with iron inclusions weighing approximately 0.5 g as specimen, which he mixed with a casting plastic and formed it into a cylindrical shape weighing 1.1 g. Aluminum interacts strongly with the electrons due to its large magnetic moment of its nuclei and is therefore well-suited for nuclear polarization. Furthermore, it is paramagnetic and responds to the magnetic moment without retaining magnetic properties. Alzofon used dynamic nuclear polarization to achieve the orientation of the aluminum nuclei. The sample was placed within a resonant cavity and a constant magnetic field of about 0.33 T generated by an electromagnet was applied. This caused the electrons of the aluminum sample to process at their Larmor frequency of about 9.5 GHz. The sample was then excited by a pulsed microwave field by an x-band microwave system with the same frequency as the Larmor frequency and a magnetic field vector perpendicular to the constant magnetic field. The crossed fields should cause the predicted effect as the electrons process even faster and orientate the nuclei. Switching the microwave off should have started a reorientation of the nuclei's magnetic moments causing a weight reduction due to the assumed effect. The pulse duration of the microwave field was 6 ms, equivalent to the relaxation time of the aluminum nuclei at room temperature, as the experiment was also conducted at room temperature. The change in weight was measured by a Mettler high-accuracy scale with a 0.01 mg resolution. The intervals of the microwave pulses are shown in Fig. 1 and accordingly the change in weight over time is shown in Fig. 2. Both figures are recreated according to the reference [2].

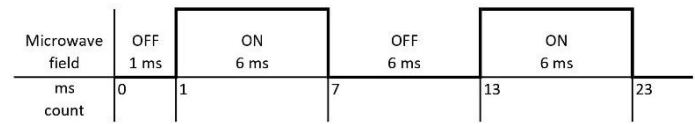

Fig. 1    Switching intervals of the microwave signal, recreated according to Alzofon's description [2, p.299]

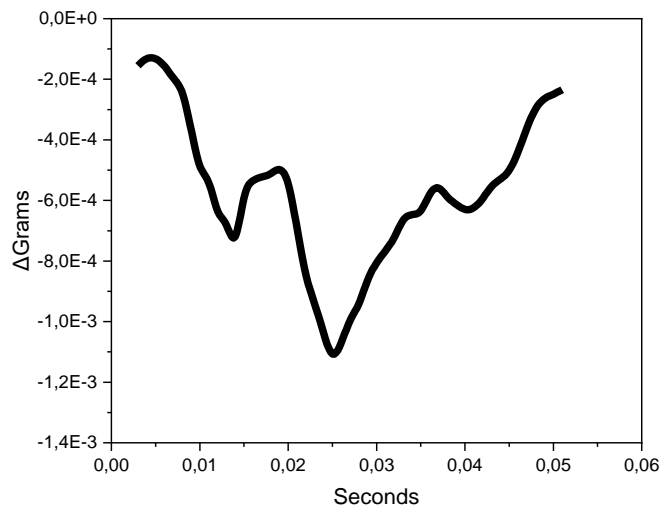

Fig. 2    Apparent measured weight change during an experiment (AF0003) by Alzofon on May 26, 1994, recreated according to reference [2, p. 300]

A maximum weight change of 1.1 mg (-1.1E-3 grams) was recorded during excitation of the pulsed microwave field. Regarding the weight ratio of aluminum, this results in a relative weight change of 0.22%. Alzofon was certain to have demonstrated the correlation between pulsed dynamic nuclear polarization and weight alteration with his experimental results.

**Experimental Setup**

In order to investigate Alzofon's claim, an experiment was carried out in cooperation with the research group for NMR spectroscopy of the Leibniz Institute for Solid State and Materials Research Dresden (IFW) to investigate the influence of nuclear magnetic resonance (NMR) polarization on the weight of test objects. For this purpose, the IFW provided one of its cryostats with a superconducting magnet generating a magnetic field strength of about 9 T. The idea was to place a sample inside the strong magnetic field within the cryostat and use NMR to investigate Alzofon's assumption. The magnetic field inside the cryostat caused the atomic nuclei of the sample to align either parallel or antiparallel to the field direction. A radiofrequency pulse was then applied to the sample. This frequency corresponds to the resonance frequency of the atomic nuclei in the magnetic field, which depends on the strength of the external magnetic field. As a result, some of the atomic nuclei were displaced from their initial alignment and brought into an intermediate position. After the pulse was turned off, the nuclei gradually returned to their original alignment according to the material-specific relaxation time. It has been investigated whether such disorientation also leads to a weakening of the gravitational field around the sample, as proposed by Alzofon. The samples were attached to a small bending beam to detect changes in the pulling force of the specimen and indicate a change in the gravitational force. The experimental setup can be attached to a 1.5 m long rod and inserted into the cryostat, see Fig. 3.

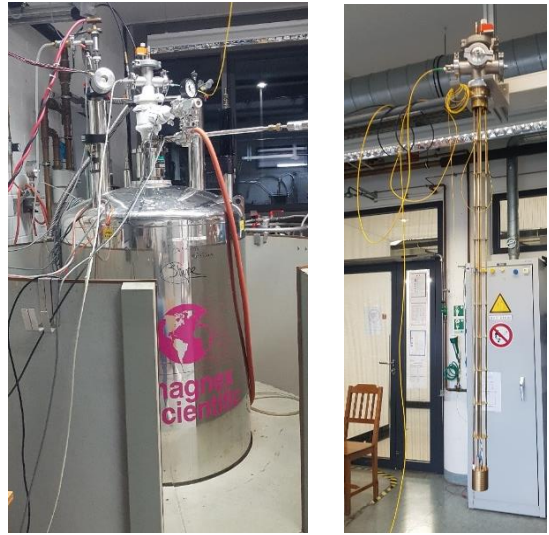

Fig. 3 Left: Cryostat with a superconducting magnet of 9 T
Right: Suspended test rod

The actual setup was mounted on the lower end of the rod. A thin aluminum strip was clamped on one end into a bracket. This strip acted as the bending beam to which various specimens could be attached. If the gravitational field around the specimen would be weakened, the pulling force of the specimen would decrease and change the deflection of the bending beam. A sensor head of an attocube IDS laser interferometer (LIF) was therefore positioned above the aluminum strip, which measured the vertical deflection of the beam. The bracket could be tilted via a linkage with bevel gear, which allowed the aluminum strip to be aligned perpendicular to the LIF. The aluminum strip was 2 mm wide and 0.1 mm thick and the free length was 18 mm, with the specimen holding point at 17 mm from the bracket. The specimens were small spheres from different materials with different diameters and were hung from the bending beam with a fine nylon thread. To displace the atomic nuclei, the specimen must be exerted to a radiofrequency pulse. Therefore, the specimen was placed inside a coil of an oscillating circuit. Fig. 4 shows a sketch of the oscillating circuit.

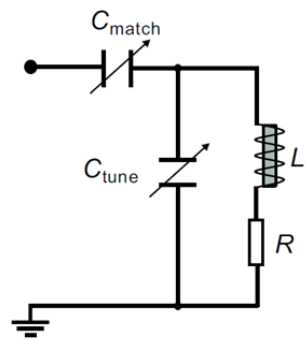

Fig. 4 Sketch of the oscillating circuit

$C_{match}$ is the matching capacitor which sets the depth of the resonance whereas $C_{tune}$ is the tuning capacitor which is used to set the resonant frequency of the oscillating circuit and $L$ is the coil in which the specimen is placed. The frequency signal itself was generated by a Tecmag Apollo NMR spectrometer. The setup at the end of the test rod can be covered with a cap to shield it from surrounding influences, such as the inflowing, cooling helium. There was no room to mount the temperature sensor on the sample, but it was placed near the coil, which was close enough to measure the temperature changes of the experiment within this cap. Fig. 5 shows a sketch of the experimental setup and the setup installed at the end of the rod.

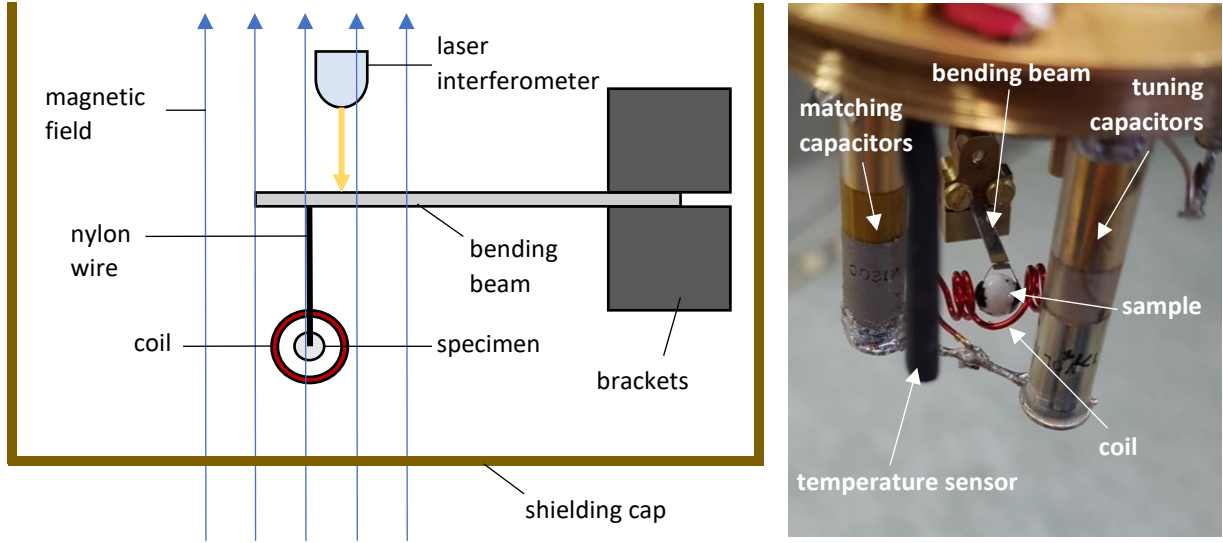

Fig. 5  Left: Sketch of the experimental setup
Right: Experimental setup at the lower end of the rod

The aluminum bending beam was as a measuring tool to detect weight changes of the specimens. The pulling force of the specimen causes a deflection of the beam which can be measured by the LIF. A change in weight would cause a change in the pulling force of the specimen and lead to a different deflection of the beam. It was therefore necessary to know the conversion factor *k* between deflection and change in weight. Fig. 6 shows a graphical representation of the deflection curve of a cantilevered beam, as it was also the case in this experiment and Table 1 shows the parameters of the aluminum bending beam.

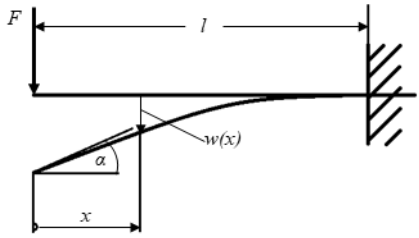

Fig. 6  Graphical representation of the deflection curve of a cantilever beam

Table 1 Parameters of the bending beam

| Parameters of the bending beam | Value |
|---|---|
| Length $l$ | 17 mm |
| Measurement point of the LIF $x$ | 4.7 mm |
| Width $b$ | 2 mm |
| Height $h$ | 0.1 mm |
| Moment of inertia $I_y$ | $\frac{b\,h^3}{12}$ |
| Young's modulus $E$ | 70 GPa [3] |

The deflection curve of a cantilever beam, as used in this experiment, is calculated as follows [4]:

$$w(x) = \frac{F\,l^3}{6\,E\,I_y}\left[2 - 3\frac{x}{l} + \left(\frac{x}{l}\right)^3\right] \tag{2}$$

This results in the following equation for the conversion factor (CF) *k*:

$$k = \frac{w(x)}{F} = \frac{l^3}{6\,E\,I_y}\left[2 - 3\frac{x}{l} + \left(\frac{x}{l}\right)^3\right] = 0.0836\,\frac{m}{N} \sim 0.82\,\frac{\mu m}{mg} \tag{3}$$

Therefore, a weight change of 1 mg would result in a measured deflection of 0.82 µm by the LIF. To confirm this experimentally, several calibration measurements at room temperature were performed with the aluminum bending beam. Precisely known weights were put on a pan that hung on the bending beam and the corresponding deflection was determined. Fig. 7 shows a graph of the determined deflections caused by various weights on the bending beam.

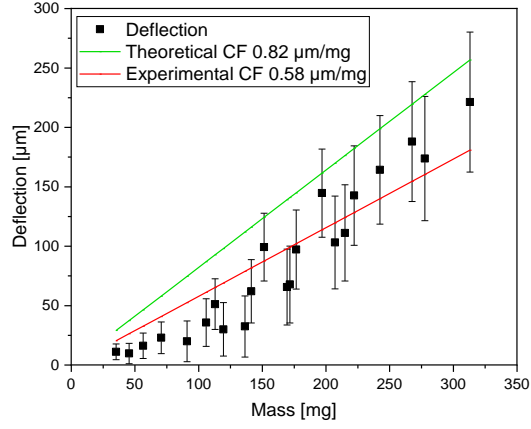

Fig. 7   Calibration measurement in comparison to the conversion factor

It can be seen, that the measured deflections are lower than estimated by the theoretical CF $k = 0.82$ µm/mg. Apparently, the aluminum bending beam used in this experiment behaves differently than expected. By alternately adding and removing weights on the bending beam, it was challenging to establish a consistent zero position, leading to drifts in the measured deflection. The values obtained from the calibration measurements yield a CF of $k = 0.58$ µm/mg, which is assumed as the worst-case assumption, and applies to experiments at room temperature. The majority of the experiments took place at very low temperatures around 5 K, where the properties of the bending beam change. The advantage of aluminum is that its Young's modulus only increases by 10% at 5 K [5]. Therefore, for measurements at this temperature, an adjusted CF of $k = 0.53$ µm/mg was applied. Since the length change of the bending beam at 5 K is less than 1%, its influence on the CF was neglected.

Not all atoms are suitable for NMR, but only atoms that have a non-zero nuclear spin and, therefore, a magnetic moment in their ground state [6]. Like in Alzofon's experiment, aluminum was chosen as a specimen. The spheres had a diameter of 5 mm with a purity of 99.75%. The most common isotope of aluminum, 27Al, has a nuclear spin of 5/2 and is therefore an excellent material for use in NMR [6]. The high nuclear spin also allows for high sensitivity of NMR spectroscopy, which is advantageous for analyzing samples at low concentrations. In addition, we also investigated brass as another metallic specimen containing 63.85% of copper. Copper has two stable isotopes, 63Cu and 65Cu, both of which have nuclear spins of 3/2 and are therefore also a suitable material for use in NMR [6]. Due to the skin effect, the radiofrequency pulse penetrates the surface of metals only to a certain depth, known as the skin depth $\delta$ [7], which can be calculated as follows:

$$\delta = \sqrt{\frac{2\,\rho_R}{\omega\,\mu}} \tag{4}$$

where $\rho$ is the electrical resistivity of the conductor, $\omega$ is the angular frequency and $\mu$ is permeability of the conductor. Simplified, for our two metallic samples of aluminum and brass, it can be assumed that $\mu \approx \mu_0$. This allows us to express Equation ( 4 ) as:

$$\delta \approx \sqrt{\frac{\rho_R}{\pi\,f\,\mu_0}} \tag{5}$$

where $f$ is the frequency of the radiofrequency pulse (in this case, the Larmor frequency of the sample) and $\mu_0$ is the magnetic constant. Table 2 shows the corresponding skin depth for the metallic samples of aluminum and brass.

Table 2  Electrical resistivity and resulting skin depth for aluminum and brass excited by radiofrequency pulses of their Larmor frequencies.

| Material | $\rho_R$ [Ω m] | $f$ [MHz] | $\delta$ [µm] |
|---|---|---|---|
| Aluminum | $2.6 \cdot 10^{-8}$ | 78.285 | 9.2 |
| Brass | $7.0 \cdot 10^{-8}$ | 79.7 | 14.9 |

The skin depth of the excitation pulse was determined to be only 14.9 µm for brass and 9.2 µm for aluminum, which means that nuclear spin polarization can be manipulated only within a thin layer on the outside of the pure metal samples. Therefore, in addition to pure Aluminum, we also investigated other materials including dielectrics, like Aluminum oxide, where the pulses penetrate the whole volume and not only the surface of the samples. According to the chemical formula of aluminum oxide $Al_2O_3$, it consists of two aluminum atoms and three oxygen atoms. The atomic weight of aluminum is 26.98 g/mol and that of oxygen is 16 g/mol [8]. It can therefore be determined that aluminum accounts for approximately 52.9% of the total weight of aluminum oxide.

The other chosen materials are either composed of copper or fluorine atoms because of their suitable polarization properties like aluminum. Yttrium barium copper oxide (YBCO) was chosen, which is a high-temperature superconductor, to excite its copper atoms. The chemical formula is $YBa_2Cu_3O_{7-x}$ and copper makes up approximately 28.6% of its weight [10, 11]. Since YBCO is superconducting, the relevant parameter for the penetration of the radio frequency pulse is not the skin depth $\delta$, but the London penetration depth of a superconductor, $\lambda$. The $B_1$ excitation field in our experiments can be calculated from the applied power, $P$, the quality factor, $Q$, of the resonant circuit, the Larmor frequency, $\omega$, and the volume of the coil, $V$, as [10]:

$$B_1 = \sqrt{\frac{\mu_0\, Q\, P}{2\, \omega\, V}} \quad (6)$$

where $\mu_0$ is the vacuum permeability. $B_1$ is calculated to be well below 100 Gauss, which is much smaller than the lower critical field $H_{c1}$ of YBCO, which is about 200 – 300 Gauss [11]. Therefore, the relevant parameter for YBCO is the penetration depth, $\lambda$, which is about 1500 Angstrom = 0.15 µm at 0 K [12].

Fluorine occurs only in a single stable isotope, 19F, which has a nuclear spin of 1/2 [6]. Specimen made of polytetrafluoroethylene (PTFE) have been investigated to excite the fluorine atoms. The chemical formula is $[C_2F_4]_n$ and fluorine makes up approximately 76% of its weight [8]. The specimens were all spheres with diameters between 5 mm, except YBCO had a diameter of 7 mm. Table 3 gives an overview of the specimens and states information about the materials diameters, Larmor frequency of the excited atoms and the spin–lattice relaxation time T1.

Table 3  Overview of the specimens

| Material | Diameter | Larmor frequency (atom) | T1 time at 5 K |
|---|---|---|---|
| Aluminum | 5 mm | 78.285 MHz (Al) | 368 ms |
| $Al_2O_3$ | 5 mm | 78.185 MHz (Al) | 39.2 s |
| Brass | 5 mm | 79.7 MHz (Cu) | 400 ms |
| YBCO | 7 mm | 79.7 MHz (Cu) | 1.1 s [13] |
| PTFE | 5 mm | 282.196 MHz (F) | 13.8 s |

The Larmor frequency of the specimens and the T1 time were determined in advance as shown in the appendix. The T1 time characterizes the rate at which the orientation of the nuclei decays to disorder. It increases with decreasing temperature, therefore, the experiments are conducted at very low temperatures around 5 K. Table 3 shows that the atoms within a purely metallic compound have a short T1 time of some milliseconds, whereas dielectric compounds have a higher T1 time of some seconds. This means, that the spin polarization decays slowly within dielectrics. This could indicate that Alzofon's assumed effect of a gravitational force reduction may be weaker in dielectrics because it lasts longer. The Larmor frequency of the material was determined in advance of the experiment and the oscillating circuit was tuned to it. Usually, a copper coil was used for the oscillating circuit.

For the experiments with brass and YBCO, a coil made of silver was used, so that only the atoms of the specimen are excited and not the coil itself.

After installing the setup on the rod and aligning the bending beam to the LIF, the rod was inserted into the cryostat to be cooled down to about 5 K by liquid helium. In order to polarize the nuclear spins, the test object was excited by pulses in the radio frequency band (30 MHz to 300 MHz) from the coil. The pulses were generated by the Tecmag Apollo NMR Spectrometer and different parameters could be set: the duration of an individual pulse, the interval between two consecutive pulses and the number of pulses.

## Measurements

The first measurements initially aimed to be close to the original experiment. Like in Alzofon's work, the sample material was aluminum - in this case, a sphere with a diameter of 5 mm - and the experiment was conducted at room temperature. The sample was excited by pulses with a frequency of 78.285 MHz (Larmor frequency of aluminum), and for each measurement, the sample was subjected to 100 pulses. The duration of each pulse was 8 µs, and the pulses were spaced 50 ms apart. The configuration for this setup-A is presented in Table 4 together with information about all the other configurations, like the tested sample and the parameters of the pulse excitation, which will be discussed later. Fig. 8 shows an example for a setup-A measurement at room temperature and a corresponding reference measurement without pulse excitation.

Table 4  Setup characteristics for the pulse excitation for various samples

| Setup | A | B | C | D | E | F | G |
|---|---|---|---|---|---|---|---|
| **Material** | 5 mm Al | 5 mm Al | 5 mm $Al_2O_3$ | 5 mm PTFE | 7 mm YBCO | 5 mm Al | 7 mm YBCO |
| **Mass** | 0.1748 g | 0.1748 g | 0.2614 g | 0.1424 g | 0.7596 g | 0.1748 g | 0.7596 g |
| **Temperature** | RT | 5 K | 5 K | 5 K | 5 K | 5 K | 5 K |
| **Frequency** | 78.285 MHz | 78.285 MHz | 78.16 MHz | 282.196 MHz | 79.7 MHz | 78.285 MHz | 79.7 MHz |
| **Number of pulses** | 100 | 100 | 50 | 50 | 50 | 1000 | 1000 |
| **Pulse duration** | 8 µs | 8 µs | 4 µs | 4 µs | 8 µs | 8 µs | 24 µs |
| **Pulse interval** | 50 ms | 50 ms | 100 ms | 100 ms | 100 ms | 5 ms | 5 ms |

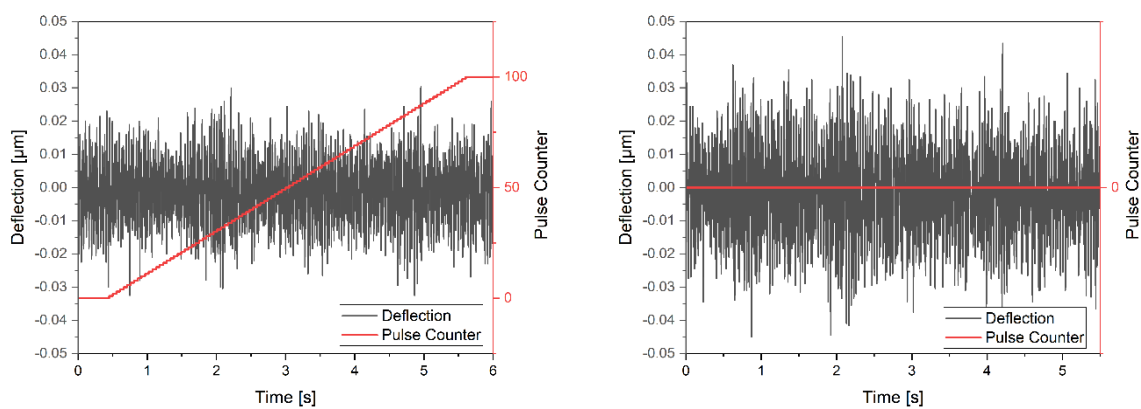

Fig. 8  Left: Excitation of a 5 mm Al sphere with 100 pulses at 78.285 MHz at room temperature (setup-A)
        Right: Reference measurement for a 5 mm Al sphere at room temperature

The measurements at room temperature showed noise in the range of +/- 0.03 µm and no measurable deflection of the bending beam was detected. A possible effect could have been covered by the noise of the measurement.

Therefore, the following measurements were carried out at 5 K (setup-B) to reduce temperature noise and furthermore to increase the T1 time for the reorientation of the nuclear spins and extent the visibility of the potential effect. The measurement was repeated with the same pulse parameters and examples for the graphs are shown in Fig. 9.

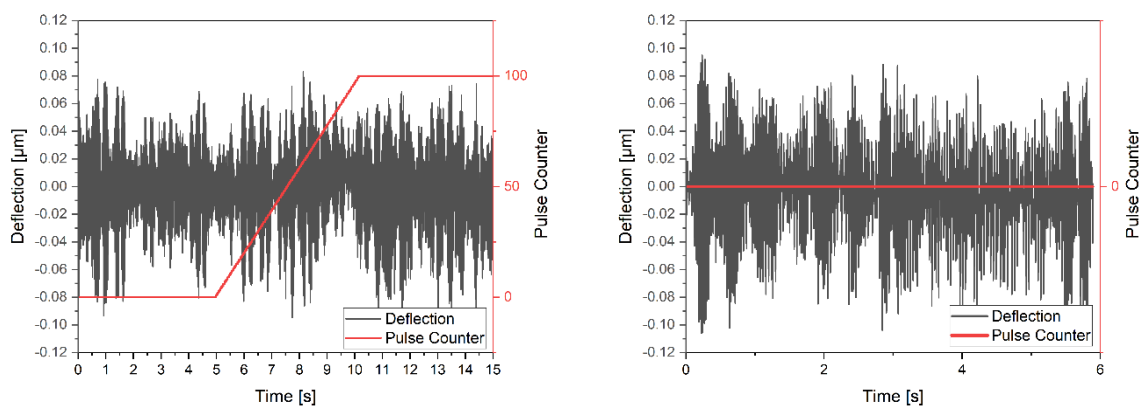

Fig. 9  Left: Excitation of a 5 mm Al sphere with 100 pulses at 78.285 MHz at 5 K (setup-B)
Right: Reference measurement for a 5 mm Al sphere at 5 K

No clear signal was observed with these measurements either, and although the temperature noise should have decreased, the noise in the low-temperature measurement is even a bit higher than at room temperature and in the range of about +/- 0.08 µm, most likely due to noise from the laboratory. Additionally, pure metallic samples, like aluminum, are generally not suitable for these measurements due to the skin effect. In the case of aluminum, the skin depth was 9.2 µm and the radiofrequency pulse only excited atomic nuclei within this thin layer.

As dielectric specimens are fully excited by the NMR pulses, further measurements were initially carried out with aluminum oxide. The parameters for the excitation (setup-C) are shown in Table 4 and the graphs are shown in Fig. 10.

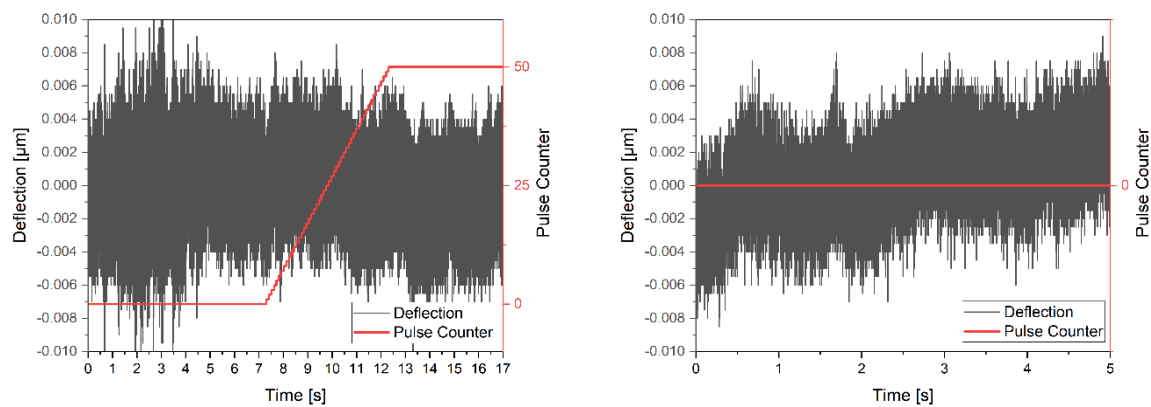

Fig. 10  Left: Excitation of a 5 mm $Al_2O_3$ sphere with 50 pulses at 78.16 MHz at 5 K (setup-C)
Right: Reference measurement for a 5 mm $Al_2O_3$ sphere at 5 K

Compared to the reference measurement, no significant deflections due to the pulse excitation were observed. In both measurements, the background noise was approximately +/- 0.005 µm, which indicates a much better resolution compared to the measurement with the aluminum sample. Converted by the low-temperature conversion factor, CF(5K), of k = 0.53 µm/mg, this corresponds to a weight resolution of about 9.4 µg.

The two other dielectric samples made of PTFE (setup-D) and YBCO (setup-E) also did not show an apparent weight change due to pulse excitation. The parameters for the excitation are shown in Table 4 and the graphs are shown in Fig. 11.

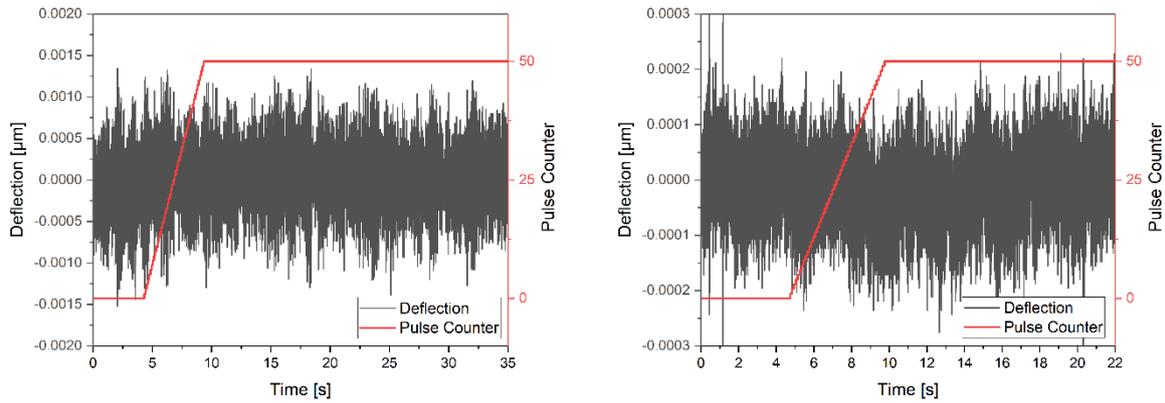

Fig. 11  Left: Excitation of a 5 mm PTFE sphere with 50 pulses at 282.196 MHz at 5 K (setup-D)
Right: Excitation of a 7 mm YBCO sphere with 50 pulses at 79.7 MHz at 5 K (setup-E)

It can be seen, that the noise has been further reduced to about 0.001 µm for PTFE and even 0.0002 µm for YBCO, which corresponds to a weight resolution of approximately 1.9 and 0.38 µg, respectively. Still, no deflection caused by nuclear polarization is visible.

As the predicted effect by Alzofon could not be reproduced with these measurements, the influence of the pulse parameters on the results was further investigated. It was found that increasing the pulse duration and setting the pulse numbers to approximately 1,000 or more, a deflection of the bending beam could be indeed observed. The graphs for measurements with aluminum (setup-F) and YBCO (setup-G) are shown in Fig. 12 and the parameters for the setups are presented in Table 4.

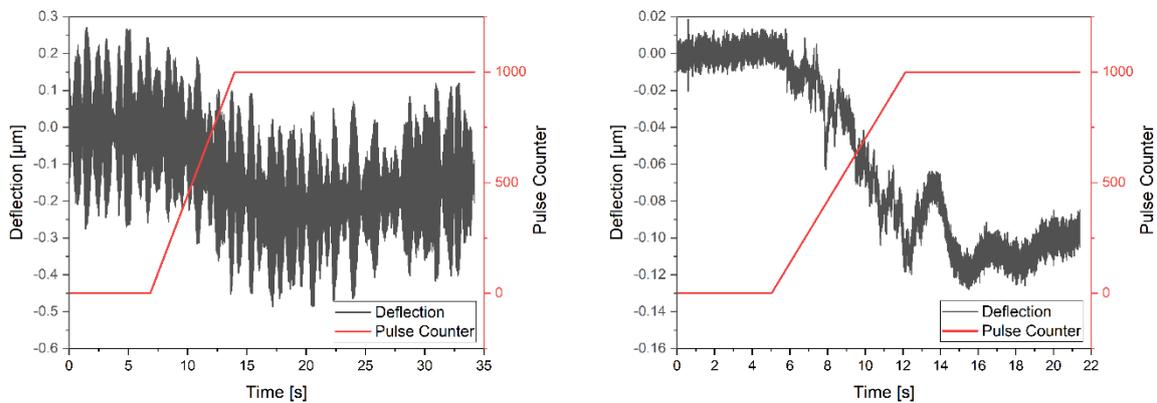

Fig. 12  Left: Excitation of a 5 mm Al sphere with 1000 pulses at 78.285 MHz at 5 K (setup-F)
Right: Excitation of a 7 mm YBCO sphere with 1000 pulses at 79.7 MHz at 5 K (setup-G)

Despite the noise of +/- 0.25 µm, the measurement with aluminum shows a slight but distinct deflection of the bending beam of about -0.25 µm, which corresponds to an upward deflection, which could be related to an apparent mass loss. However, this change seems to build up slowly, within about 10 s after the start of the pulses, and then apparently gradually decreases. In this case, the deflection corresponds to an mass change of approx. -0.47 mg. In the measurement with YBCO, the pulse duration was increased from 8 µs to 24 µs compared to the aluminum measurement and generally shows a significantly lower noise of about +/- 0.01 µm, with a more distinct deflection becoming visible. Approx. one second after the start of the pulses, a deflection builds up, reaching its maximum at about -0.11 µm towards the end of the pulses. This corresponds to an approx. mass change of -0.21 mg. These measurements show similarities with the measurements of Alzofon, who measured mass changes around 1.1 mg, see Fig. 2. Due to the short measurement time, no statements can be made about the decay behavior of the deflection for this measurement.

In order to investigate the origin of these deflections, the parameters of the measurements were changed systematically. It was found that increasing pulse duration and decreasing the pulse interval makes the deflection peak more distinct and increases the amplitude of the peaks. A distinct deflection was measured after switching on the pulses, which also returned to the normal state after the end of the pulse set. Although, the deflection of the bending beam always occurred with a time delay of about 1 s. Note, that the deflection decreases, which means that the beam bends upwards. This would indeed correlate to a lower weight or lower acting gravitational force. To make the deflections more visible, the measurement period was extended and the number of sets of pulses applied to the specimen was increased for each measurement. An example of such a measurement is shown in Fig. 13 with the pulse characteristic presented in Table 5.

Table 5  Pulse characteristics for a 5 mm Al sphere with 10 pulse sets

| Material | 5 mm Al |
|---|---|
| Temperature | 5 K |
| Frequency | 78.285 MHz |
| Number of pulse sets | 10 |
| Time between each set | 30 s |
| Number of pulses per set | 50 |
| Pulse duration | 32 µs |
| Pulse interval | 5 ms |

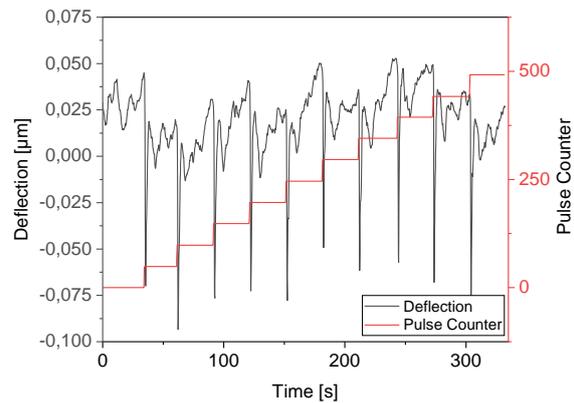

Fig. 13  Aluminum specimen 5 mm with 50 pulses per set at 78.285 MHz

The specimen is an aluminum sphere with a diameter of 5 mm, which was excited by a pulse frequency of 78.285 MHz at 5 K. 50 pulses with a duration of 32 µs and a spacing of 5 ms formed a set of pulses. For each measurement, the sample was exposed to 10 such sets of pulses with a time interval of 30 s between each set. The graph shows a clear transient deflection at the end of each pulse set with an amplitude of approx. -0.1 µm. This corresponds to a transient apparent change in weight of approx. -0.19 mg.

In the following, the number of pulses has also been increased and it was found that the deflection of the beam increases as well. Exciting the specimen to over 10,000 pulses per set, the deflection of the beam reaches a steady state. The stable displacement also decreased within a certain time after switching off the pulses. Fig. 14 and Fig. 15 show the increasing displacement when increasing the number of pulses. The specimen was a brass sphere with a diameter of 5 mm, which was exposed to several sets of pulses with a frequency of 79.7 MHz at 5 K. The number of pulses per set changed with every measurement and accounted for 100; 500; 1,500 and 20,000 pulses. The measurement with 100 pulses shows a particularly strong similarity to the measurements of Alzofon (Fig. 2).

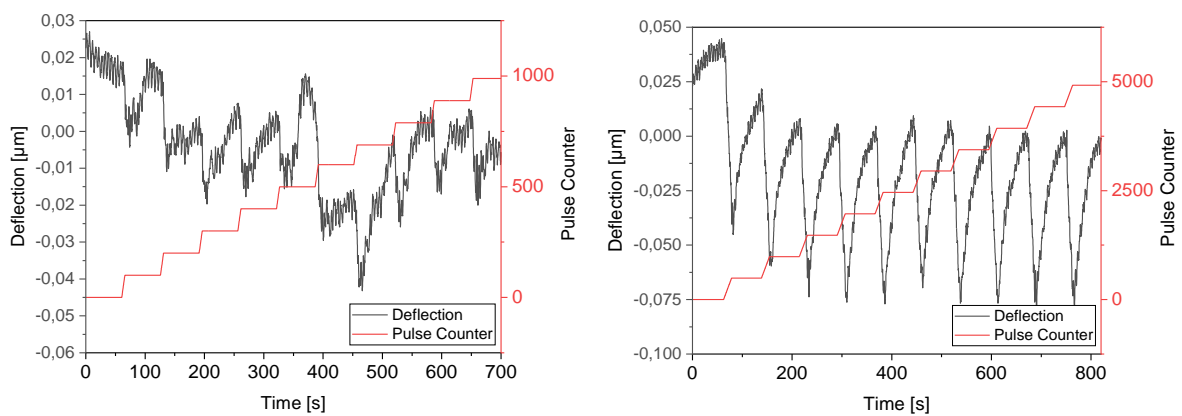

Fig. 14  Left: Brass specimen 5 mm with 100 pulses per set at 79.7 MHz
        Right: Brass specimen 5 mm with 500 pulses per set at 79.7 MHz

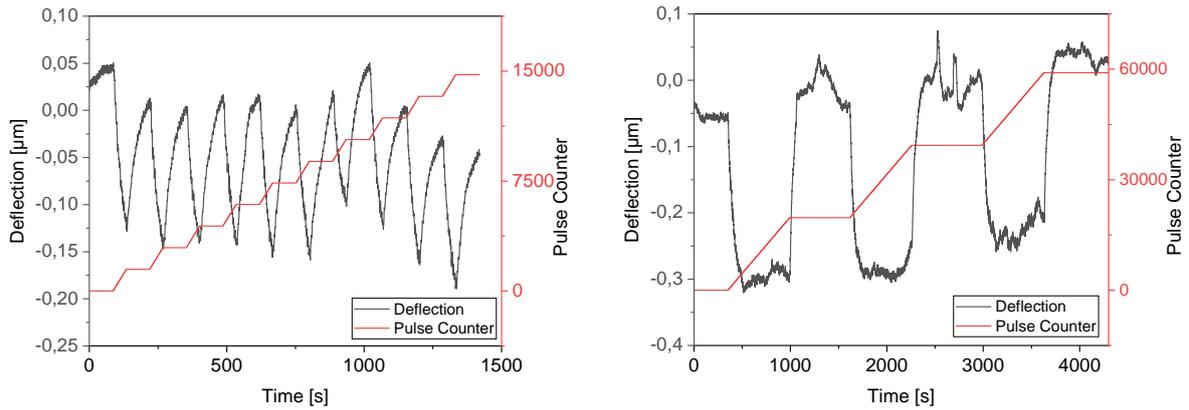

Fig. 15  Left: Brass specimen 5 mm with 1,500 pulses per set at 79.7 MHz
         Right: Brass specimen 5 mm with 20,000 pulses per set at 79.7 MHz

It can be seen how the deflection of the bending beam increases with the number of pulses until a maximum deflection, a kind of equilibrium state, is reached. This state remains stable over the time of excitation until the pulses are shutdown. Subsequently, the deflection almost returns to the initial state. In the case of brass, for example, the maximum deflection was approx. -0.25 µm, which correlates to an apparent weight change of -0.47 mg.

Numerous measurements have been conducted to investigate the influence of the number of pulses, the pulse interval and the pulse duration. Various samples and parameters were examined to link the deflection of the beam to a cause. Fig. 16 shows the influences of these single parameters on the displacement of the bending beam.

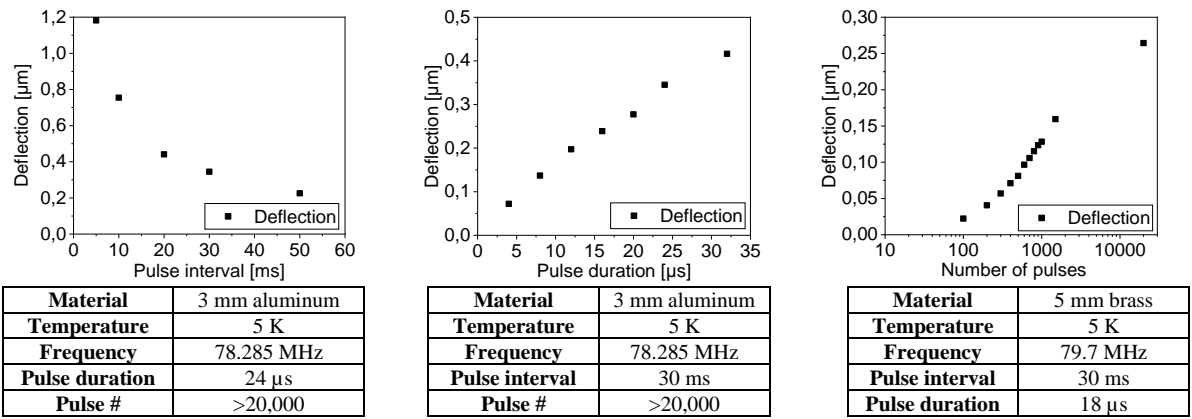

| Material | 3 mm aluminum |
|---|---|
| Temperature | 5 K |
| Frequency | 78.285 MHz |
| Pulse duration | 24 µs |
| Pulse # | >20,000 |

| Material | 3 mm aluminum |
|---|---|
| Temperature | 5 K |
| Frequency | 78.285 MHz |
| Pulse interval | 30 ms |
| Pulse # | >20,000 |

| Material | 5 mm brass |
|---|---|
| Temperature | 5 K |
| Frequency | 79.7 MHz |
| Pulse interval | 30 ms |
| Pulse duration | 18 µs |

Fig. 16  Overview of different measurement values on the displacement of the bending beam

It can be seen that the displacement of the bending beam increases under the following conditions:

- Decreasing pulse interval (time between two consecutive pulses)
- Increasing pulse duration
- Increasing the number of pulses

In the course of investigating the cause of the deflections and whether they were a material-specific effect, reference measurements without a specimen have been conducted. Deflections were also observed when the parameters were changed in favor of the conditions mentioned above. Fig. 17 and Fig. 18 show measurements without a test object, similar to those described before. Fig. 17 shows on the left a short-term measurement with 50 pulses with a frequency of 79.7 MHz at 5 K and on the right the corresponding reference measurement. A particularly low noise level of +/- 0.00005 µm is observed, which would correspond to a weight resolution of about 94 µg. Fig. 18 shows a measurement with 1,000 pulses on the left and a measurement with three sets of 10,000 pulses each on the right, also with a frequency of 79.7 MHz at 5 K.

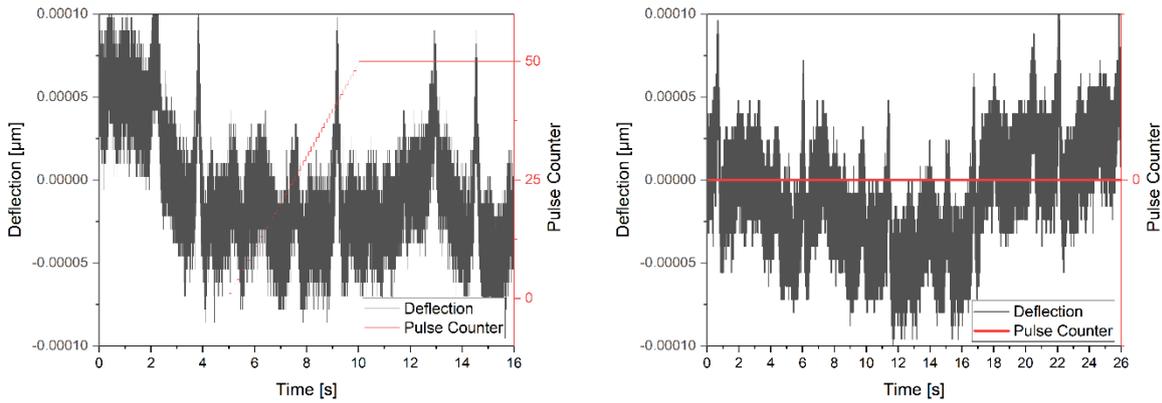

Fig. 17  Left: Excitation without a specimen with 50 pulses at 79.7 MHz at 5 K
Right: Reference measurement without a specimen at 5 K

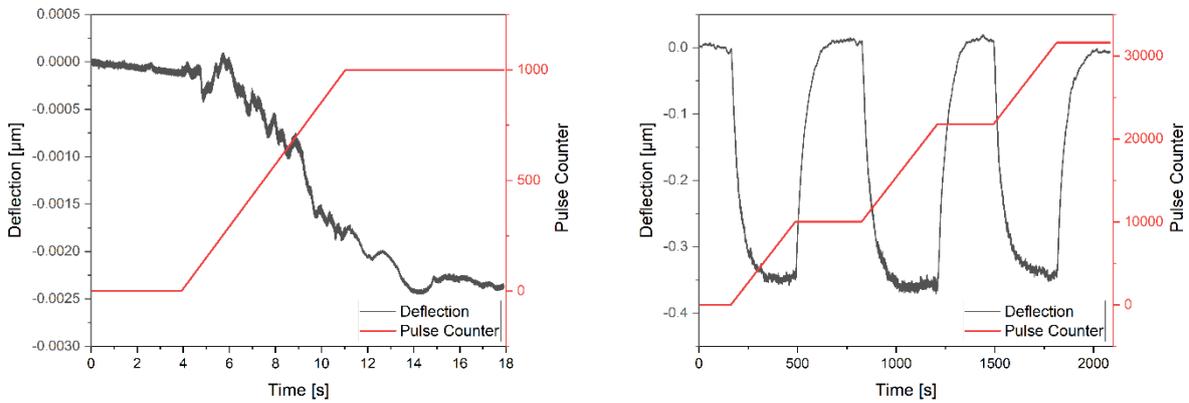

Fig. 18  Left: Excitation without a specimen with 1,000 pulses at 79.7 MHz at 5 K
Right: Excitation without a specimen to 3 sets of 10,000 pulses at 79.7 MHz at 5 K

Even without a specimen, a similar behavior is observed as in measurements with a specimen. The apparent mass change does not seem to originate from the specimen itself, but from other causes. Exciting an object with a set of pulses always comes with a certain amount of energy, depending of the parameters of the pulses. Assuming that the energy density, or power, of such a pulse set is described as the energy per time, all of the three conditions named above (decreasing pulse interval, increasing pulse duration and increasing the number of pulses), increase the amount of power of a pulse set. And with power, one has to consider heat and the temperature of the experiment. Looking at the temperature graphs for the individual measurements, a clear similarity in their distribution can be observed. Fig. 19 shows the deflection and the temperature during a measurement with and without a specimen attached to the bending beam. Note, that the temperature scale is inverted.

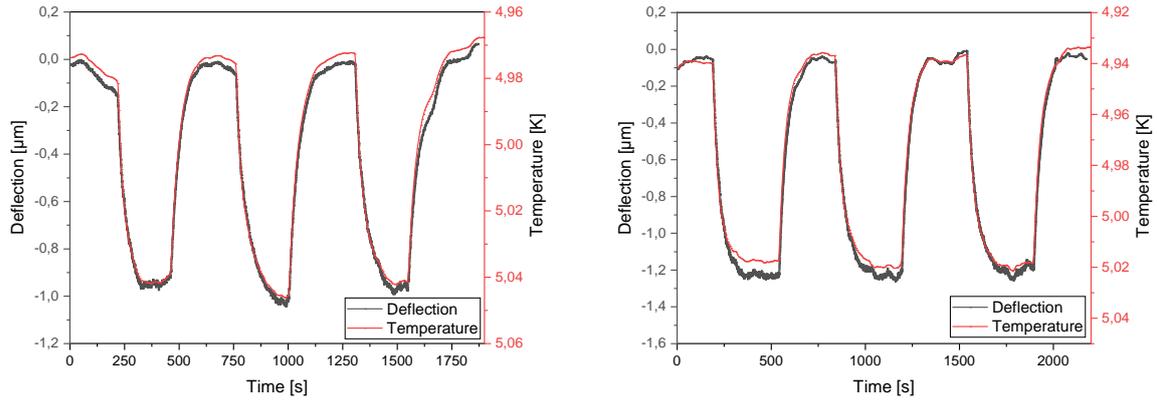

Fig. 19  Left: Aluminum specimen 5 mm with 20,000 pulses per set at 78.285 MHz
        Right: No specimen with 20,000 pulses per set at 78.285 MHz

The graphs show that curves of displacement and temperature are almost aligned. Once the temperature in the vicinity of the experiment increases, a negative deflection is measured, and vice versa. This indicates that the measured deflection of the beam is directly connected to the temperature change in the environment of the experiment.

**Discussion**

Alzofon used dynamic nuclear polarization (DNP) to experimentally proof his hypothesis and claimed to have measured changes in mass of a test body. In this experiment, it was investigated whether the claimed effect also occurs using nuclear magnetic resonance spectroscopy. One difference between both methods is the proportion of polarized atomic nuclei, which is at least ten times higher using DNP. In NMR spectroscopy, the expected value of nuclear spin polarization $\langle \hat{I}_z \rangle$ is calculated as follows [10]:

$$\langle \hat{I}_z \rangle = \frac{\gamma \, h_P \, I \, (I+1) \, H_0}{3 \, k_B T} \qquad (7)$$

where $\gamma$ is the gyromagnetic ratio of the nucleus, $h$ is the Planck constant, $I$ is the nuclear spin, $H_0$ is the external magnetic field strength, $k_B$ is the Boltzmann constant and $T$ is the temperature. In the case of this experiment, the external magnetic field $H_0$ was 9 T and the temperature T was 5 K. Table 6 shows the gyromagnetic ratio of the nucleus of the investigated nuclei and the resulting expected value of nuclear spin polarization at 5 K.

Table 6 Expected value of nuclear spin polarization of the investigated nuclei

| Nuclei | $\gamma$ [MHz/T] | $\langle \hat{I}_z \rangle$ at 5K |
|---|---|---|
| $^{27}$Al | 11.094 | $2.8 \cdot 10^{-3}$ |
| $^{63}$Cu | 11.285 | $1.2 \cdot 10^{-3}$ |
| $^{19}$F | 40.059 | $8.7 \cdot 10^{-4}$ |

It can be seen that the proportion of nuclear spin polarization $\langle \hat{I}z \rangle$ is very low. Only 0.28% of the aluminum nuclei are polarized. For copper, it's only 0.12%, and for fluorine, it's as low as 0.087%. In turn, only this small fraction of polarized nuclei can generate the effect claimed by Alzofon. $\langle \hat{I}z \rangle$ only applies to those samples that were fully penetrated by the radiofrequency pulse. In this experiment these were the two samples made of $Al_2O_3$ and PTFE. Due to skin effects, the radiofrequency pulse penetrates the surface of pure metallic specimens only to a certain depth. Table 2 shows that the skin depth is 9.2 µm for aluminum and 14.9 µm for brass. Considering that the samples each have a diameter of 5 mm, a percentage of excited volume $\eta$ is obtained. The same can be done for YBCO, in which case the relevant parameter for the penetration of the radio frequency pulse is not the skin depth $\delta$, but the London penetration depth $\lambda$ and is about 0.15 µm. Even if we assume a ten times larger penetration depth at 5 K, the radiofrequency pulse penetrates only 1.5 µm of the spherical sample, and therefore affects only a tiny fraction of the nuclear spins, even less than for a normal metal. With a diameter of 7 mm for the YBCO specimen, the percentage of excited volume $\eta$ can also be calculated. From this percentage $\eta$, only $\langle \hat{I}_z \rangle$ portions are subjected to nuclear spin polarization. Consequently, this results in a nuclear spin polarization ratio $\Pi$ for the entire sample body from both factors. These values for aluminum, brass and YBCO are shown in Table 7.

Table 7  Portion of actual nuclear spin polarization $\Pi$ due to the skin effect for aluminum and brass.

| Material | $\eta$ | $\langle \hat{I}_z \rangle$ at 5K | $\Pi$ |
|---|---|---|---|
| Aluminum | 0.011 | $2.8 \cdot 10^{-3}$ | $3.1 \cdot 10^{-5}$ |
| Brass | 0.018 | $1.2 \cdot 10^{-3}$ | $2.2 \cdot 10^{-5}$ |
| YBCO | 0.0013 | $1.2 \cdot 10^{-3}$ | $1.5 \cdot 10^{-6}$ |

The aluminum sphere had a mass of 0.1748 g and a purity of 99.75%. With $\Pi$ from Table 7, the nuclear spin polarization of aluminum nuclei accounts for 5.4 µg of the entire aluminum sample. The brass sphere had a mass of 0.5527 g and a copper ratio of 63.85%. Consequently, with $\Pi$ from Table 7, the nuclear spin polarization of copper nuclei accounted for 7.4 µg of the entire copper sample. For the $Al_2O_3$ sphere, according to $\langle \hat{I}_z \rangle$ from Table 6 for aluminum and a mass of 0.2614 g with an aluminum content of 52.9%, the nuclear spin polarization of aluminum nuclei was 387.2 µg in the entire $Al_2O_3$ sample. The PTFE sphere had a mass of 0.1424 g and a fluorine content of approximately 76%. Therefore, with $\langle \hat{I}_z \rangle$ for fluorine, the nuclear spin polarization of fluorine nuclei was 94.2 µg in the PTFE sample. For YBCO with a specimen mass of 0.7596 g and containing approximately 28.6% copper, the affected mass for nuclear spin polarization by the factor $\Pi$ accounts for 0.34 µg.

The measurements mostly showed very low noise in the range of 0.0002 µm to 0.01 µm, which would correspond to a weight resolution of 0.38 µg to 18.9 µg. Only for aluminum, the noise was higher, ranging from 0.03 µm to 0.25 µm, which would correspond to a weight resolution of 56.6 µg to 472 µg. For measurements without a test object, the noise was reduced to only 0.05 nm, approx. 94 ng. Table 8 shows the affected mass by the nuclear spin polarization for the different samples and the minimum noise of the measurement converted into an equivalent change in mass according to the conversion factor $k$ from Equation ( 3 ).

Table 8  Overview of mass affected by nuclear spin polarization (NSP mass) of the investigated sample materials and their measurement noise converted into an equivalent change in mass.

| Material | NSP mass | Min. noise |
|---|---|---|
| Aluminum | 5.41 µg | 56.6 µg |
| Brass | 7.41 µg | 8.5 µg |
| $Al_2O_3$ | 387 µg | 9.4 µg |
| PTFE | 94.2 µg | 1.9 µg |
| YBCO | 0.34 µg | 0.38 µg |

The metallic samples, aluminum and brass, have a small mass fraction affected by the NSP due to the skin effect. For aluminum, this is only 5.4 µg, and in addition, the minimal noise was 56.6 µg. Consequently, no statements can be made for the measurements with aluminum. For brass, the mass affected by the NSP was 7.4 µg and therefore only slightly below the minimal noise of 8.5 µg. These results also do not allow for a conclusion. $Al_2O_3$ and PTFE have a much higher NSP mass due to the complete penetration of the radiofrequency pulse. For $Al_2O_3$, it is 387.2 µg compared to a minimum noise of 9.4 µg. PTFE has a NSP mass of 94.2 µg with noise at 1.9 µg. The measurements with YBCO showed very low noise of 0.38 µg but therefore the sample also shows a very low penetration depth. Even if we generously overestimate it to be 1.5µm at 5 K, the NSP mass accounts only 0.34 µg and is below the minimum noise of the measurements. Since only a few pulses are required for the NSP and to generate the effect claimed by Alzofon, even at a low number of pulses, a deflection should be measurable. However, Al2O3 (Fig. 10) and PTFE (Fig. 11 left) show no deflection by such pulse excitation. Perhaps the effect claimed by Alzofon is significantly smaller in dielectrics, as the nuclei reorientate over a much longer T1 time compared to metallic specimens. Additionally, the NMR technique itself could be a reason why the Alzofon effect cannot be confirmed or eventually the effect is significantly smaller than the measurement noise.

However, with an increased number of pulses, significant deflections were measured in every material. Like all electrical circuits, the oscillating circuit of the test setup generates heat loss, especially when increasing the power of the frequency pulse sets. The surrounding helium heats up and expands, changing its density and also its refractive index. Since the displacement is measured by a laser interferometer, this change in the refractive index creates an apparent deflection of the beam, which explains why deflections of the bending beam could be measured even without any specimen attached to the bending beam. Therefore, it is necessary to consider the temperature changes and associated changes in refractive index during the measurements in the evaluation. The LIF operates at a wavelength of 1.53 µm. The refractive index of helium at 0 °C and a pressure of 101325 Pa for this wavelength is [14]:

$$n_{He;\ 1.53\mu m;\ 0°C;\ 101325Pa} = 1{,}000034709 \tag{8}$$

For the evaluation of an influence by the temperature $T$, the Lorentz-Lorenz equation [15] has to be considered, which describes the physical relationship between refractive index $n$ and density $\rho$:

$$\frac{n^2 - 1}{\rho(n^2 + 2)} = const. \tag{9}$$

For gases with a refractive index close to one, the equation can be expressed in a simplified form as:

$$\frac{2(n-1)}{3\rho} = const. \quad (n \approx 1) \tag{10}$$

Considering the ideal gas law, where $p$ is the pressure and $R_s$ is the specific gas constant:

$$p = \rho \cdot R_s \cdot T \tag{11}$$

It is assumed that the pressure inside the cryostat remains constant. Therefore, the dependence between refractive index $n$ and temperature $T$ can be expressed as:

$$\frac{n_\infty - 1}{(n_0 - 1)} = \frac{T_0}{T_\infty} \tag{12}$$

or

$$n_\infty = \frac{T_0}{T_\infty} \cdot (n_0 - 1) + 1 \tag{13}$$

Where $T_0$ is the temperature and $n_0$ is the refractive index of the initial state; and $T_\infty$ is the temperature and $n_\infty$ is the refractive index of the altered state. During each measurement, the temperature was also measured and it was assumed that the pressure was constant at all time. The temperature during the pulse excitation, $T_2$, increased compared to the initial temperature, $T_1$ and the corresponding refractive indices, $n_1$ and $n_2$, can be calculated according to equation ( 13 ). From these the refractive error $RE$ could be determined for every displacement measurement $x$ as follows:

$$RE = x \left(\frac{1}{n_1} - \frac{1}{n_2}\right) \tag{14}$$

Taking into account, that the temperature change causes a refraction error, Fig. 20 shows the measured deflection of all measurements that have been conducted and also shows the theoretical measurement error due to the changing refractive index because of the temperature change.

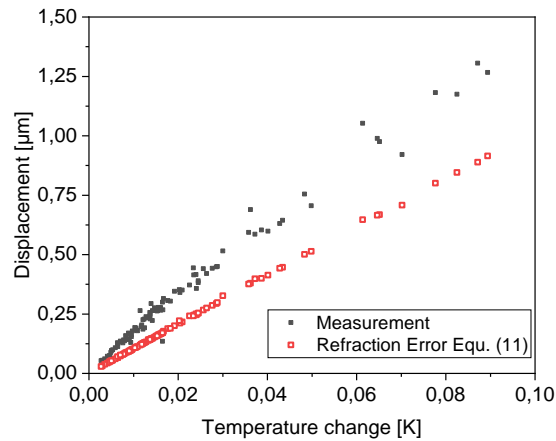

Fig. 20 All displacement measurements depending on the temperature change and the theoretical refraction error caused by this change in temperature

It can be seen that the measurement point distributions are similar. The refraction error is linearly dependent of the temperature change and the displacement measurements show the same behavior. As the temperature sensor is not located close to the measuring section between bending beam and laser interferometer above but sideways under the bending beam, the actual temperature change is probably slightly higher. This would in turn increase the refraction error. Also, a possible local pressure change along the measuring section might have increased the error. It can therefore be assumed that what appears to be a measured deflection is purely due to the change in temperature of the environment. The aluminum bending beam rests during the whole measurements while only the refractive index of the surrounding changes. This explanation also applies to the measurements mentioned at the beginning, in which the deflections were measured with a time delay, because the heat propagation is a time-dependent process and not an immediate effect. The heat generated at the coil below the bending beam warms the surrounding helium which than rises and streams through the measuring section between bending beam and laser interferometer. This can also explain temporary peaks, as short sequences of pulses heat up a small region around the coil and this "cloud" of higher temperature rises, streams through the measurement path, and thus can be briefly measured as a deflection.

## Conclusion

Alzofon rose attention with his concept about influencing the gravitational field of objects. According to his theory, the gravitational influence on an object is reduced as it goes through an inner process from an ordered system to a disordered system. Therefore, a series of experiments were carried out to create such an effect using NMR spectroscopy and nuclear spin polarization to detect transient weight changes. Spherical specimens made from aluminum, brass, $Al_2O_3$, PTFE and YBCO were hung from a small bending beam inside a cryostat and the orientations of the nuclear spins of the atoms of a specimen were aligned by a very strong magnetic field. Pulse excitation with its Larmor frequency cause a fraction of the aligned nuclear spins to be flipped and gradually align again, which should cause a possible change in weight according to Alzofon. This change in weight could be detected using a laser interferometer to measure the deflection of the bending beam.

The initial measurements to verify Alzofon's idea were conducted with low numbers of pulses, which would still be sufficient to orientate the nuclei and cause the predicted effect. However, no measurable weight change was observed, even though resolutions up to 0.38 µg were achieved. When the pulse numbers and duration were increased, deflections of the bending beam were measured, which indicated such an apparent change in weight. In order to find out whether these deflections origin from a possible change in the gravitational field or are an artifact of some other influence, numerous measurements have been conducted with different parameters for the pulse sequences and different materials. All tested materials responded to the excitation at their Larmor frequency. It was also shown that the deflection of the bending beam increased with increasing power of the signal; either by increasing the pulse duration or increasing the number of pulses or reducing the pulse interval. Given the increased power of the signal, the possibility of unintentional heat input arose. It was found that a portion of the pulse energy was transferred to the surrounding helium, causing it to heat up and change its refractive index. The changing refractive index in the LIF measurement path resulted in an apparent measurement deflection, especially in measurements with large pulse numbers or long pulse duration. All measured deflections turned out to be a temperature artifact.

In addition, it was shown that an effect on the order of milligrams, as described by Alzofon, could not be detected in this experiment. The small penetration depth of the excitation pulses in metals and YBCO, only a few micrometers, may have greatly weakened the effect. In the case of dielectrics, the significantly longer T1 time could be the reason why the effect of gravitational field attenuation appears to be weakened due to the extended relaxation time. The measurements of this experiment showed resolutions in the range of micrograms and suggest that the effect in question is either not measurable using NMR spectroscopy, smaller than predicted or non-existent.

## Acknowledgement

We gratefully acknowledge the helpful discussions with George Hathaway.

**Appendix**

**A1**

Fig. A 1 shows a frequency analysis for specimens in which aluminum nuclei were supposed to be excited. For pure metallic aluminum, the Larmor frequency is 78.285 MHz. For $Al_2O_3$, the Larmor frequency is 78.185 MHz.

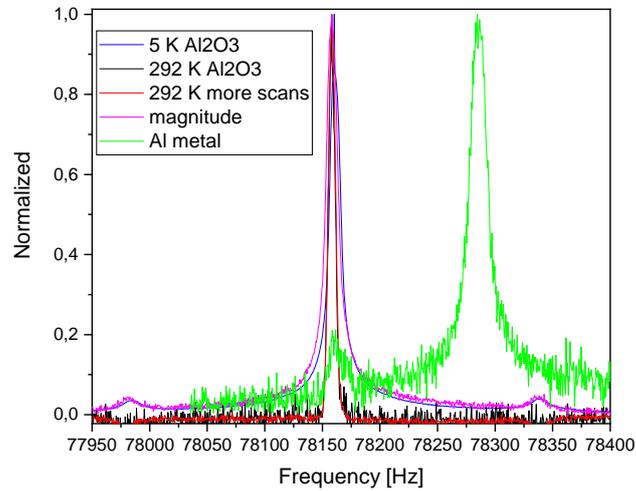

Fig. A 1 Frequency analysis for aluminum and $Al_2O_3$

**A2**

Fig. A 2 shows the measurement data for determining the T1 times of aluminum at room temperature and at 5 K. Aluminum has a T1 time of 6.71 ms at room temperature and 367.7 ms at 5 K.

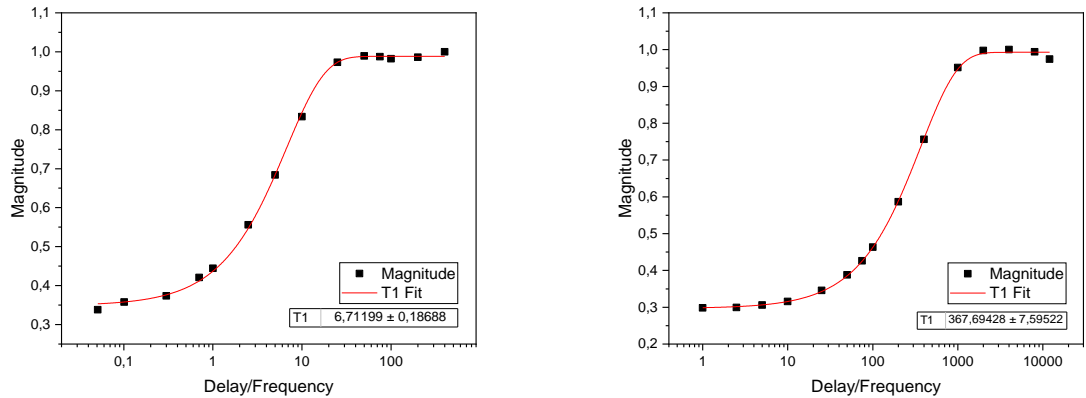

Fig. A 2 Left: Determination of T1 time for aluminum at room temperature T1 = 6.71 ms
　　　　Right: Determination of T1 time for aluminum at 5 K T1 = 367.7 ms

**A3**

Fig. A 3 shows the measurement data for determining the T1 times of $Al_2O_3$ at room temperature and at 5 K. $Al_2O_3$ has a T1 time of 11.9 s at room temperature and 39.2 s at 5 K.

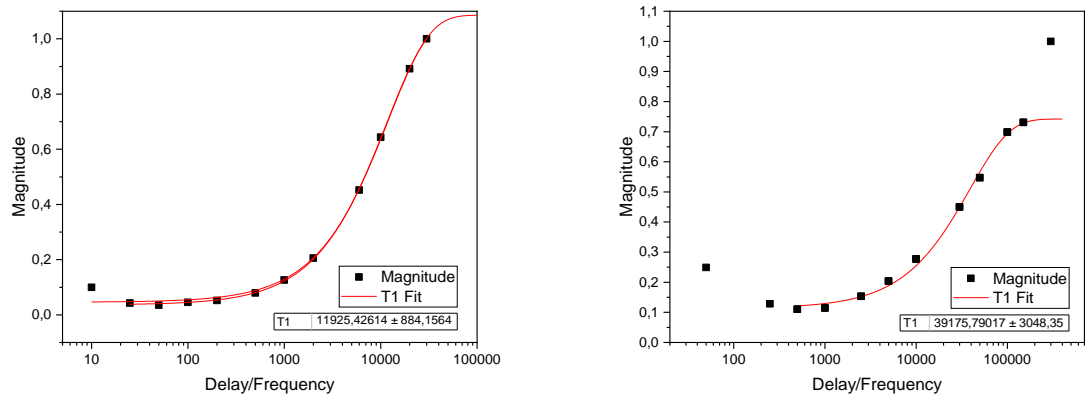

Fig. A 3 Left: Determination of T1 time for $Al_2O_3$ at room temperature T1 = 11.9 s
Right: Determination of T1 time for $Al_2O_3$ at 5 K T1 = 39.2 s

**A4**

Fig. A 4 shows a frequency analysis for the PTFE specimen in which fluor nuclei were supposed to be excited. The Larmor frequency for PTFE at 4.5 K is 282.196 MHz.

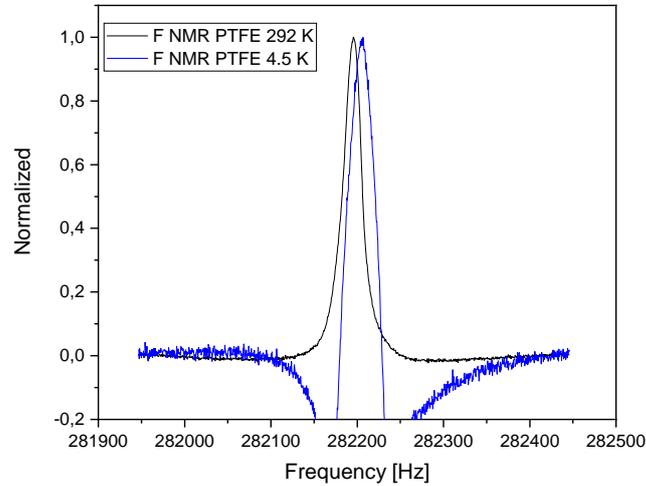

Fig. A 4 Frequency analysis for PTFE

**A5**

Fig. A 5 shows the measurement data for determining the T1 times of PTFE at room temperature and at 5 K. PTFE has a T1 time of 0.659 s at room temperature and 13.8 s at 5 K.

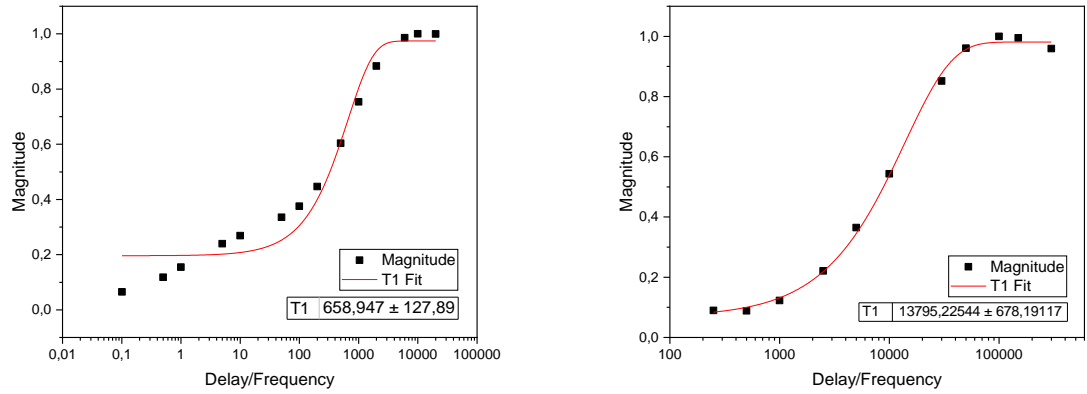

Fig. A 5 Left: Determination of T1 time for PTFE at room temperature T1 = 0.659 s
Right: Determination of T1 time for PTFE at 5 K T1 = 13.8 s